\documentclass{aa}

\usepackage{graphicx}
\usepackage{txfonts}
\usepackage[dvipsnames]{xcolor}
\usepackage{xparse,xcolor}

\begin{document}

   \title{Dwarf nova outbursts in intermediate polars}

   \author{J.-M. Hameury
          \inst{1,2}
          \and
          J.-P. Lasota\inst{3,4,2}
          }

   \institute{Observatoire Astronomique de Strasbourg, Université de Strasbourg, CNRS UMR 7550, 67000 Strasbourg, France\\
             \email{jean-marie.hameury@astro.unistra.fr}
         \and
         Kavli Institute for Theoretical Physics, Kohn Hall, University of California, Santa Barbara, CA 93106, USA
         \and
         Institut d'Astrophysique de Paris, CNRS et Sorbonne Universit\'es, UPMC Paris~06, UMR 7095, 98bis Bd Arago, 75014 Paris, France
\and
             Nicolaus Copernicus Astronomical Center, Polish Academy of Sciences, Bartycka 18, 00-716 Warsaw, Poland               
}

   \date{}


  \abstract
  {The disc instability model (DIM) has been very successful in explaining the dwarf nova outbursts observed in cataclysmic variables. When, as in intermediate polars (IP), the accreting white dwarf is magnetized, the disc is truncated at the magnetospheric radius, but for mass-transfer rates corresponding to the thermal-viscous instability such systems should still exhibit dwarf-nova outbursts. Yet, the majority of intermediate polars in which the magnetic field is not large enough to completely disrupt the accretion disc, seem to be stable, and the rare observed outbursts, in particular in systems with long orbital periods, are much shorter than normal dwarf-nova outbursts.}
   {We investigate the predictions of the disc instability model for intermediate polars in order to determine which of the observed properties of these systems can be explained by the DIM.}
   {We use our numerical code for the time evolution of accretion discs, modified to include the effects of the magnetic field, with constant or variable mass transfer from the secondary star.}
   {We show that intermediate polars have mass transfer low enough and magnetic fields large enough to keep the accretion disc stable on the cold equilibrium branch. We show that the infrequent and short outbursts observed in long period systems, such as e.g., TV Col, cannot be attributed to the thermal-viscous instability of the accretion disc, but instead have to be triggered by an enhanced mass-transfer from the secondary, or, more likely, by some instability coupling the white dwarf magnetic field with that generated by the magnetorotational instability operating in the accretion disc. Longer outbursts (a few days) could result from the disc instability.}
   {}


   \maketitle
%

\begin{table*}
\caption{Observed properties of outbursting IPs} 
\label{tab}
\centering 
\begin{tabular}{l l l l l l l } 
\hline\hline 
Name & $P_{\rm orb}$ & $\Omega_{\rm spin}$ & duration &  $\tau_{rec}$ & comments & Refs. \\
     & (hr)          & (mHz)               & (days)   &         &          &       \\ 
\hline 
V455 And & 1.35 & 92.9 & 30 days & & WZ Sge type superoutburst & 1 \\
CC Scl & 1.40 & 16.1 & 9 (SO) & & SU UMa type & 2 \\
HT Cam & 1.43 & 12.2 & 1-2 & & & 3 \\
DW Cnc & 1.44 & 2.71 & 2-3  & & & 4 \\
EX Hya & 1.64 & 1.56 & 2-3  & 1.5 yr & & 5\\
V1223 Sgr & 3.36 & 8.43 & 0.25-1 & & & 6 \\
YY Dra (= DO Dra) & 3.96 & 11.9 & 3-5  & 1 yr & & 7 \\
TV Col & 5.49 & 3.29 & 0.25 & & & 8, 9\\
XY Ari & 6.61 & 30.5 & 5 &  & & 10 \\
CXOGBS J174954.5-294335 & 8.61 & 12.5 & $> 0.25$ & & & 11\\
GK Per & 47.9 & 17.9 & 60 & 2-3 yr & historical nova & 12 \\
\hline
\end{tabular}
\tablebib{
(1)~\citet{nhz09}; (2)~\citet{kho15}; (3)~\citet{i02}; (4)~\citet{cbg08}; (5)~\citet{hkn00}; (6)~\citet{vv89}; (7)~\citet{ach08}; (8)~\citet{sm84}; (9)~\citet{hss05}; (10)~\citet{hmb97}; (11)~\citet{jth17}; (12)~\citet{s15} 
}
\end{table*}

\section{Introduction}

Dwarf novae (DNe) are cataclysmic variables (CVs), i.e. binary systems consisting of a white dwarf accreting from a low-mass companion, that undergo recurrent outbursts. These outbursts usually last for a few days, and their recurrence time is of order of weeks to months \citep[see e.g.][for a review]{w95}. There is overwhelming evidence that these outbursts are well explained by the thermal-viscous disc instability model (DIM). In this model, the viscosity is parametrized according to the $\alpha$ prescription \citep{ss73}; the disc becomes thermally and viscously unstable when its temperatures is of the order of the ionisation temperature of hydrogen and the opacities are a very sensitive function of temperature \cite[see][for a review of the DIM]{l01}. At any given radius, one can determine the surface density $\Sigma$ of the disc as a function of the local mass transfer rate (or equivalently the effective temperature $T_{\rm eff}$), and the $\Sigma - T_{\rm eff}$ curve has an S-shape with the hot upper branch and cool lower branches being stable, and the intermediate one unstable. The DIM accounts well for the observed properties of dwarf novae: recurrence times, outburst durations and amplitudes, provided that some additional ingredients are included in the model, most notably a change in the viscosity parameter $\alpha$ between the hot and cool branches. Some other effects must be taken into account such as for example the truncation of the accretion disc in quiescence as a result of the white dwarf magnetic field or of disc evaporation close to the white dwarf; one must also include the possibility of enhanced mass transfer from the secondary \citep{hlw00}, even though it is unclear that irradiation of the secondary can lead to a significant increase of the mass loss rate because the $L_1$ Lagrangian point is shielded by the accretion disc \citep{vh07,vh08}. 

According to the DIM, a system is stable if the disc temperature is everywhere high enough that hydrogen is fully ionized, which occurs for high mass transfer rates, or low enough that hydrogen is neutral or in molecular form, which can happen, if the disc extends all the way down the white dwarf surface, for vanishingly small mass transfer rates. Observations are in agreement with this prediction; claims that \object{SS Cyg} would be at a distance of 166 pc \citep{hms99} would have invalidated the DIM \citep{sh02, ls07}, and the conclusion that the distance to SS Cyg must indeed be of order of 100 pc \citep{ls07} was later confirmed by radio observations \citep{msk13} and by the parallax as determined by Gaia \citep{gaia}, which set the source distance at 116 pc.

As mentioned earlier, the magnetic field of the white dwarf has a significant influence on the stability properties of the accretion disc, since, for low mass-transfer rates, the outbursts starts in the innermost parts of disc that are cut off by magnetic truncation. This is the reason why the presence of a magnetic field has been invoked in some dwarf nova systems; this field is usually assumed to be strong enough to truncate the disc in quiescence, but weak enough that it does not channels the accretion onto the magnetic poles during outbursts, which would cause pulsations at the white dwarf magnetic spin period; such pulsations have not been observed. Such magnetic field have been first suggested to account for the very long recurrence time of systems such as \object{WZ Sge} \citep{lhh95,wlt96}\footnote{\citet{wlt96} could not, however, account for the length of WZ Sge (super)outburst.} and later used in a more general context.    

There are, however, systems in which the white dwarf is observed to be strongly magnetized. In the AM Her stars (or polars), the magnetic field is so strong that the rotation of the white dwarf is synchronous with the orbit, and matter flows directly from secondary star onto the white dwarf; these systems do not contain accretion discs. In the so-called intermediate polar systems (IPs), the white dwarf is not synchronous with the orbit; depending on the field strength, some of these systems may not possess an accretion disc \citep{hkl86}, but it is clear that the vast majority of these systems do possess an accretion disc. These discs should then be subject to the same thermal-viscous instability as in non-magnetic or weakly magnetic systems, and indeed some of the IPs do show recurrent outbursts. These are, however, different from the standard dwarf nova outbursts; their recurrence time is much longer, and their duration can be as short as a few hours. Surprisingly, little work has been devoted to applying the DIM to IPs; \citet{av89} suggested that the observed outburst duration and recurrence time of these systems are well explained if the disc is truncated at some large radius. This early work, however, relies on several disputable assumptions; in particular the inner disc radius is supposed to be the same in quiescence and in outburst, the viscosity is higher ($\alpha = 1$) than is now deduced from observations \citep[$\alpha \approx 0.2$,][]{Smak99,KL12} is used on the hot branch of the S curve) and the disc thermodynamics is treated in a simplified way. \citet{kwm92} considered the case of \object{GK Per} that is a long period (2 days) CV showing long and infrequent outbursts, and concentrated on the shape and duration of these outbursts.

In this paper, we perform a systematic study of the DIM when the white dwarf magnetic field is large. We briefly summarise the observational properties of IP exhibiting outbursts in Section 2; in Section 3, we apply the DIM to magnetic systems. We conclude that the DIM correctly predicts that many systems should be stable on either the cold or the hot branch, except for systems with short orbital period which could undergo normal dwarf nova outbursts. The DIM cannot, however, explain the very short outbursts (duration less than 1 day) observed in long period systems. These could be due to a mass transfer instability that we consider in section 4; or, more speculatively, to the interaction of the magnetic field generated by the magnetorotational instability (MRI) with the white dwarf magnetic field.

   \begin{figure}
   \centering
   \includegraphics[width=\columnwidth]{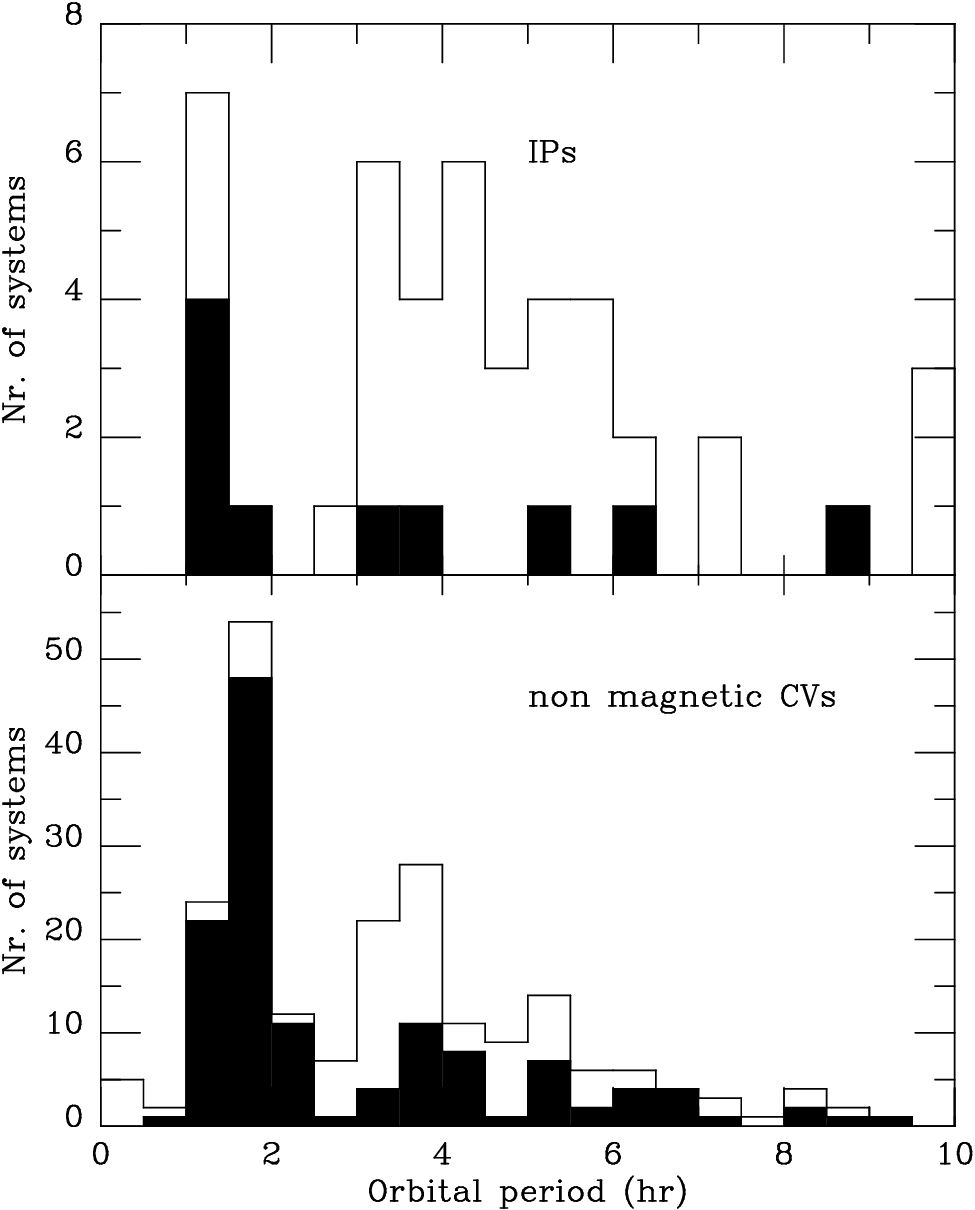}
   \caption{Period distribution of outbursting (black) magnetic and non magnetic CVs, as compared to the total population}
   \label{fig:ips}
   \end{figure}
   
\section{Observational properties of IPs and outbursts}

Figure \ref{fig:ips} shows the orbital period distribution of outbursting and steady IPs, as compared to that of non-magnetic CVs. Data are taken from Koji Mukai's on-line catalogue\footnote{https://asd.gsfc.nasa.gov/Koji.Mukai/iphome/iphome.html} for the IPs; we have selected only the confirmed IPs for which the orbital period is well determined. For the non-magnetic CVs, we ave used the updated \citet{rk03} catalogue\footnote{http://wwwmpa.mpa-garching.mpg.de/RKcat/}. The list of outbursting IPs is given in Table \ref{tab}.

It clearly appears from Fig. \ref{fig:ips} that there is a strong deficit of outbursting IPs, both below and above the 2-3 hr period gap. At short orbital periods, all non magnetic systems show outbursts, with very few exceptions due to either uncertainties in the orbital period, as for HV And \citep{rr12}, uncertainties in the classification (SX LMi and BZ Cir, not classified as DNe in \citet{rk03} catalogue have shown outbursts, see \citealt{wth98,kiu09}), or to the fact that the white dwarf underwent a recent nova explosion (T Pyx and BK Lyn, the latter system having transited in 2005 to a dwarf nova, \citealt{puk13}). CP Pup, classified as a nova-like non magnetic CV, is probably an IP \citep{mom13}. The 3 out of 8 IPs with periods below 2 hr which do not exhibit dwarf nova outbursts are therefore a very significant exception.

Similarly, there is a very significant excess of stable long period IPs. For non-magnetic CVs, the fraction of outbursting systems is 42\%. If this ratio were the same in IPs, one would have expected to have 16 IPs classified as dwarf novae, whereas only 5 are found.

Not only are the outbursts rare in IPs, they are also much shorter than in non-magnetic systems with comparable orbital periods, in particular at long orbital periods; they can in some cases (TV Col, CXOGBS J174954.5-294335) last for less than one orbital period. It is therefore plausible that, as will be discussed below, the very short outbursts are not caused by the thermal/viscous instability of the DIM, but call for a different explanation; this would make the deficit of DNe among IPs even more significant.

\section{Disc instabilities}

\begin{table*}
\caption{Model parameters and results} 
\label{tab:results}
\centering 
\begin{tabular}{l l l l l l l l l l l l l} 
\hline\hline 
Nr & $P_{\rm orb}$ & $q$ & $\mu_{33}$ & $\alpha_{\rm c}$ & $\alpha_{\rm h}$ & $\dot{M}_{\rm tr}$ & $\dot{M}_{\rm max}$\tablefootmark{a} & $ \log \dot{M}_{\rm min}$\tablefootmark{b} & $\tau_{\rm rec}$\tablefootmark{c} & duration\tablefootmark{d} & $\Omega_{\rm K} (r_{\rm mag})$ & $\nu\Sigma$\tablefootmark{e}\\ 
         & (hr)        &  &            &                 &      &$(10^{16} \rm g/s)$&$(10^{16} \rm g/s)$ & g/s & (days) & (days) & (mHz) \\   
\hline 
1 & 6 & 0.6 &0 & 0.02 & 0.1 & 10 & 59 & 12.25 & 62.2 & 9.70 & -- & 0 \\
2 & 6 & 0.6 &0.01 & 0.02 & 0.1 & 10 & 71 & 14.76 & 87.8 & 12.9 & 43.0 & 0\\
3 & 6 & 0.6 &0.1 & 0.02 & 0.1 & 10 & 100 / 87 & 15.56 & 197 / 95 & 15.8 / 4.8 & 13.2 & 0\\
4 & 6 & 0.6 &0.1 & 0.04 & 0.2 & 10 & 88 & 15,61 & 54.1 & 6.2 & 13.8 & 0\\
5 & 6 & 0.6 &0.1 & 0.01 & 0.1 & 10 & 182 / 153 & 15,40 & 346.8 / 173.4 & 13.7 / 4.7 & 11.2 & 0\\
6 & 6 & 0.6 &0.1 & 0.02 & 0.2 & 10 & 184 / 157 & 15.45 & 196 / 98 & 8.0 / 2.5 & 11.8 & 0 \\
6 & 6 & 0.6 &0.1 & 0.02 & 0.1 & 10 & 59.8 & 15.76 & 83.2 & 14.2 & 16.0 & $\dot{M}/3\pi$ \\ 
6 & 6 & 0.6 &0.1 & 0.02 & 0.1 & 2 & 66,9 / 29,6 & 15.42 & 678 / 333 & 15.8 / 3.7 & 11.5 & $\dot{M}/3\pi$  \\
7 & 6 & 0.6 &0  & 0.02 & 0.1 & 2 & 86 / 1.6 & 12.25  &  1217 / 56.5 &  15.2 / 1.9  & -- & 0\\
8 & 6 & 0.6 &0.01 & 0.02 & 0.1 & 2 & 84 / 37 & 14,76 & 1269 / 254 & 15.5 / 6.4 & 43.0 & 0 \\
9 & 1.5 & 0.2 &0  & 0.02 & 0.1 & 5 & 5.2 & 12,25 &  60.4   &   5.8   & -- & 0 \\
10 & 1.5 & 0.2 &0.01 & 0.02 & 0.1 & 5 & 12 & 14,72 &  377.5  &   10.1 & 41.3 & 0\\
11 & 1.5 & 0.2 &0.01 & 0.02 & 0.1 & 5 & 4.7 & 14.76 & 92.2 & 9.1 & 43.0 & $\dot{M}/3\pi$  \\
\hline
\end{tabular}
\tablefoot{
When there is a sequence of large and smaller outbursts, the first number refer to the properties of the largest outbursts, the second to the smallest ones.\\
\tablefoottext{a}{Peak accretion rate during outbursts.}
\tablefoottext{b}{Minimum mass accretion rate during quiescence.}
\tablefoottext{c}{When there is  sequence of large and smaller outbursts, the first number is the time interval between large outbursts, the second the minimum time interval between consecutive outbursts.}
\tablefoottext{d}{Outburst duration.}
\tablefoottext{e}{Boundary condition at inner disc edge.}
}
\end{table*}

   \begin{figure}
   \centering
   \includegraphics[angle=-90,width=\columnwidth]{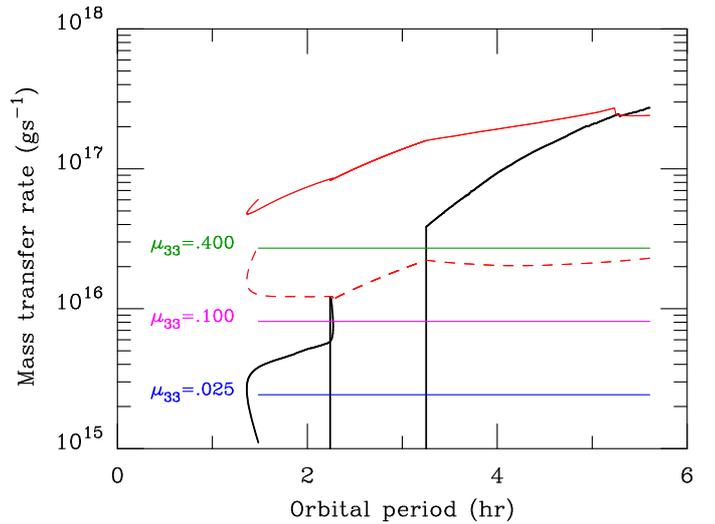}
   \caption{Critical mass transfer for stability. The upper red curve represents the critical mass transfer rate above which discs are hot and stable. The expected  secular mean of the mass transfer is given by the black curve (see text for details). The green, magenta and blue curves represent the critical mass transfer below which discs are cold and stable (Eq. \ref{eq:mdotmax}), for magnetic moments of $4 \times 10^{32}$, $\times 10^{32}$ and $2.5 \times 10^{31}$ G cm$^3$ respectively. The disc exists only if its inner radius is smaller than the circularization radius; the dashed red curve shows the maximum possible mass transfer allowing the disc to be stable and cold at $1.5 r_{\rm circ}$.}
   \label{fig:stab}
   \end{figure}

\subsection{Stability criteria}

According to the DIM, an accretion disc is hot and stable if the accretion rate $\dot{M}$ is everywhere in the disc larger than a certain critical value $\dot{M}_B$; as this critical value increases with increasing distance from the white dwarf, this is occurs when the mass transfer rate from the secondary $\dot{M}_{\rm tr}$ is larger than $\dot{M}_B (r_{\rm out})$, where $r_{\rm out}$ is the outer disc radius. When additional heating terms in the outer disc (heating due to the dissipation of tidal torques and by the stream impact on the disc edge, \citealt{bhl01}) are included, the disc is stable for mass-transfer rates smaller than $\dot{M}_B(r_{\rm out})$. From our simulations it follows that the disc is stable if
\begin{equation}
\dot{M}_{\rm tr} > \dot{M}_B (0.8 r_{\rm out}) = 9.5 \times 10^{15} \alpha^{-0.004} M_1^{-0.88} \left( \frac{r_{\rm out}}{10^{10} \rm cm} \right)^{2.65} \; \rm g s^{-1},
\end{equation}
where we have used the fit for $\dot{M}_B$ given by \citet{l01}; $M_1$ is the primary mass in solar units and $\alpha$ the Shakura-Sunyaev viscosity parameter.

The disc is also cold and stable if $\dot{M}_{\rm tr}$ is everywhere smaller than a critical value $\dot{M}_A$; as this critical value also increases with $r$, this is possible if
\begin{equation}
\dot{M}_{\rm tr} < \dot{M}_A (r_{\rm in}) = 4.0 \times 10^{15} \alpha^{-0.01} M_1^{-0.89} \left( \frac{r_{\rm in}}{10^{10} \rm cm} \right)^{2.68} \; \rm g s^{-1},
\label{eq:mdota}
\end{equation}
where $r_{\rm in}$ is the inner disc radius. If the disc is not truncated, for hydrogen-dominated discs this is possible only for extremely small, unrealistic mass-transfer rates\footnote{Cold stable helium-dominated discs are observed in the case of AM~CVn systems \citep[see e.g.,][]{Smak83,KLDH12}.}; however, the magnetic field truncates the white dwarf at the magnetospheric radius given by
\begin{equation}
r_{\rm mag} = 2.66 \times 10^{10} \mu_{33}^{4/7} M_1^{-1/7} \left( \frac{\dot{M}_{\rm acc}}{10^{16} \; \rm g s^{-1}} \right)^{-2/7} \; \rm cm,
\label{eq:rmag}
\end{equation}
where $\dot{M}_{\rm acc}$ is the accretion rate onto the white dwarf, which differs from $\dot{M}_{\rm tr}$ if the system is not steady and $\mu_{33}$ is the white dwarf magnetic moment in units of $10^{33}$ Gcm$^3$.

The usual no-stress inner-boundary condition $\nu \Sigma = 0$, where $\nu$ is the kinematic viscosity ad $\Sigma$ the surface density, has a strong stabilizing effect.  Because of this condition, the critical, maximum value of  $\Sigma$ on the cold branch cannot be reached at the inner radius, but only at some distance from it. Our simulations have shown that when the disc is not very far from being stable, inside-out outbursts are triggered at approximately at $1.5 r_{\rm in}$.  Equation (\ref{eq:mdota}) should therefore be evaluated at that position in the disc; combining this with Eq. (\ref{eq:rmag}) the stability condition then becomes
\begin{equation}
\dot{M}_{\rm tr} < 4.89 \times 10^{16} M_1^{-0.72} \mu_{33}^{0.87} \; \rm g s^{-1}.
\label{eq:mdotmax}
\end{equation}

The truncation radius cannot, however, be larger than the circularization radius at which matter would form a circular orbit with the specific angular momentum it had when leaving $L_1$:
\begin{equation}
r_{\rm circ} = 0.0859q^{-0.426} a = 3.03 \times 10^9 q^{-0.426} (M_1 + M_2)^{1/3} P_{\rm hr}^{2/3}\; \rm cm,
\label{eq:rcirc}
\end{equation}
where $a$ is the orbital separation, $q=M_2/M_1$ the mass ratio, $P_{\rm hr}$ the orbital period in hours and $M_2$ the secondary mass in solar units. Equation (\ref{eq:rcirc}) is a fit to \citet{ls75} table 2; this fit is accurate to better than 1\% for $0.06 < q < 1$. This implies that if $\dot{M}_{\rm tr} < \dot{M}_A (1.5 r_{\rm circ}$), there exists a magnetic moment for which the disc can be stable; otherwise, the system is either unstable or has no disc.

Figure \ref{fig:stab} shows these critical rates as a function of the orbital period for various values of the magnetic moment, as compared to the expected mass transfer from the secondary as estimated by the revised model of \citet{kbp11} for which the primary mass is 0.7 $M_\odot$. It clearly appears that IPs below the period gap can be stabilized on the cold branch by a moderate magnetic field, of order of $5 \times 10^{31}$ G cm$^3$, insufficient to disrupt a disc. Dwarf nova outbursts at these short orbital periods are possible if the magnetic field is smaller than that, or if the mass transfer rate is higher than the secular mean by a factor of few.  Above the period gap, systems can be stable on the cold branch if their mass transfer rate is significantly smaller than the secular mean, and if the white dwarf magnetic moment is larger than 10$^{32}$ G cm$^3$; larger magnetic fields would cause the disruption of the disc, so that the system would appear as either a polar, if it can be synchronized, or as a discless IP if it cannot. Stable discs on the hot branch are possible for large enough mass transfer rates and long enough orbital periods, as for non-magnetic CVs.

\subsection{The case of FO Aqr}

\object{FO Aqr} is an intermediate polar which has never shown outbursts, and can therefore be considered as steady, but was observed to enter a low state in 2016 (more than 2 magnitudes decrease), from which it gradually recovered between May and July 2016 \citep{lgk16}. The partial eclipse observed during the low state is narrower and shallower than in the high state, indicating that an accretion disc is still present, but that is is smaller than during the high state. Because the low state light curve is integrated over a significant duration, it is possible, however, that the disc could have entirely disappeared during a fraction of the low state, and that it reappeared only when the luminosity had significantly increased.

The fact that no outburst has been observed during the rise through the instability strip has two possible explanations: either the accretion disc is present at all times, but is always cold and stable, which implies that the mass transfer is low (few times $10^{15}$ g $^{-1}$) during the high state, close to $\dot{M}_{\rm A}(r_{\rm mag})$, and that the magnetic moment if of order of a few times $10^{32}$ G cm$^3$, or the disc vanishes during the very low state and reappears only during the rise \citep{hl02}. We leave the investigation of these two possibilities for a further paper.

   \begin{figure}
   \centering
   \includegraphics[angle=-90,width=\columnwidth]{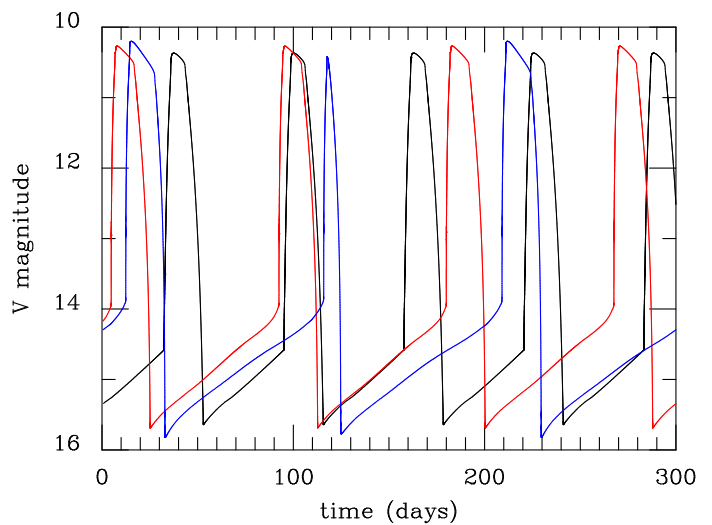}
   \caption{Time evolution of a system with a 6 hr orbital period, for magnetic moments of 0 (black curve), 10$^{31}$ (red curve) and 10$^{32}$ (blue curve) G cm$^3$.}
   \label{fig:longP}
   \end{figure}
   
   \begin{figure}
   \centering
   \includegraphics[angle=-90,width=\columnwidth]{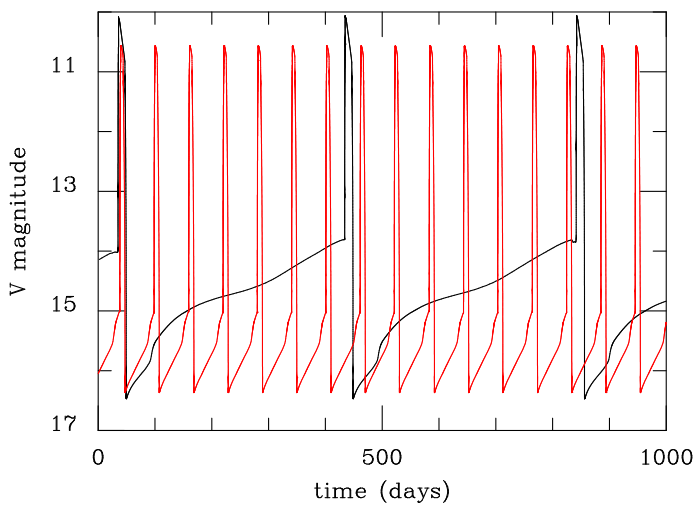}
   \caption{Time evolution of a system with a 1.5 hr orbital period, for magnetic moments of 0 (red curve) and 10$^{31}$ (black curve) G cm$^3$.}
   \label{fig:shortP}
   \end{figure}
   
\subsection{Time evolution of unstable IPs}

In order to follow the time evolution of dwarf nova outbursts in IPs, we have used the version of the DIM described in \citet{hmdlh98} in which we have incorporated heating terms due to the tidal torque and stream impact as described in \citet{bhl01}. The disc is truncated at the magnetospheric radius given by Eq. (\ref{eq:rmag}). We  considered a system with $M_1 = 0.8$M$_\odot$, $q=0.6$, and $P_{\rm orb} = 6$hr, appropriate for systems such as \object{TV Col} or \object{XY Ari}, and $\dot{M}_{\rm tr} = 10^{17}$g s$^{-1}$, slightly smaller than the expected secular mean, which ensures that the system lies in the unstable regime for dwarf nova outbursts. We further used  $\alpha_{\rm c}=0.02$, and $\alpha_{\rm h} = 0.1$. Figure \ref{fig:longP} shows the time evolution of this system. The characteristics of the light curve (recurrence time $\tau_{\rm rec}$, outburst duration, peak of the mass transfer rate and quiescent level are given in Table \ref{tab:results}. As can be seen, for the relatively high mass transfer rates used here, the impact of disc truncation is limited. On the other hand, for lower mass transfer rates ($\dot{M}_{\rm tr} = 2 \times 10^{16}$g s$^{-1}$, the effect of disc truncation is large: the recurrence time is very significantly increased, and reaches 254 days between two consecutive outbursts, and 3.5 yr between two major ones for a magnetic moment of $10^{31}$ Gcm$^3$. For $\mu$ ten times larger, we found that the system is marginally stable on the cold branch, although it should have been marginally unstable according to Eq. (\ref{eq:mdotmax}); this is because our code, due to its implicit nature, is able to follow systems on the unstable branch when they are close to the stability limit.

Figure \ref{fig:shortP} shows the evolution of a system with a 1.5 hr orbital period, for magnetic moments of 0 and $10^{31}$ Gcm$^3$. The secondary mass is taken to be 0.16 M$_\odot$, and the mass transfer rate is $5 \times 10^{15}$ g s$^{-1}$, typical for this period. Truncation of the inner disc by the white dwarf magnetic field has a major impact on outbursts: the recurrence time is increased by a factor 6, and the outbursts are longer and brighter.

Reducing $\alpha_{\rm c}$ leads to longer recurrence time while reducing $\alpha_{\rm h}$ leads to longer outbursts. Our results are in line with those of \citet{av89} as far as the recurrence times are concerned, although the value of $\alpha_{\rm c}$ we assume is significantly smaller then theirs (0.02 instead of 0.1), possibly because of the large truncation radius they assume; we never obtain outbursts as short as those of \citet{av89} because their value for $\alpha_{\rm h}$ is unrealistically high. We can find realistic parameters for the magnetic moment that can fit the observed recurrence times; we cannot, however, reproduce the very short durations observed in some systems, such as V1223 Sgr or TV Col. 

\subsection{Inner boundary condition}

We have used the standard boundary condition $\nu\Sigma =0$ at the inner disc edge, which corresponds to no torque at the inner disc's rim. This boundary condition is well adapted to discs around black holes where an innermost stable orbit exists \citep{BP00}, but might not be suitable when the accretion discs is separated from the stellar  surface by a boundary layer \citep[see e.g., Chapter 6 in][]{FKR02}. A similar situation occurs when the disc is truncated by a magnetic field. \citet{hl02} find that the inner boundary condition becomes
\begin{equation}
\nu \Sigma = \frac{\dot{M}}{3 \pi} f \left| \frac{\Omega_{\rm in} -
\Omega_*}{\Omega_{\rm in}} \right|,
\label{eq:nusig}
\end{equation}
where $f \leq 1$ is a numerical factor that accounts for possible matter ejection by the propeller effect, $\Omega_*$ is the rotation frequency of the white dwarf, and $\Omega_{\rm in}$ is the Keplerian frequency at the inner edge of the disc. Equation (\ref{eq:nusig}) is similar to the condition obtained by \citet{dt91} in a slightly different context. The surface density is increased as compared to the standard case, as well as the luminosity because of the work done by the torque, the energy source being the rotation of the white dwarf which is partly redistributed over the whole disc. The stabilizing effect of the $\nu\Sigma=0$ condition may therefore disappear. 

One should note that the accretion rate entering Eq. (\ref{eq:rmag}) is not the mass transfer rate at the inner edge of the disc, but is instead
\begin{equation}
\dot{M}_{\rm acc} = \dot{M}(r_{\rm mag}) + \pi \Sigma \frac{d}{dt} (r_{{\rm mag}}^2).
\end{equation}
The time derivative term is required for mass conservation and accounts for the fact that a fraction of the mass flow which reaches the inner edge is used to build the disc inner edge when the disc is progressing towards the white dwarf, and conversely that the white dwarf can accrete more than the incoming mass if the disc is recessing. 

The propeller effect plays a role if the Kepler orbital frequency at the inner disc edge $\Omega_{\rm K}$ is less than the spin frequency of the white dwarf $\Omega_*$, given in Table \ref{tab}. Comparing $\Omega_*$ with $\Omega_{\rm K} (r_{\rm mag})$, as given in Table \ref{tab:results} shows that the propeller will be ineffective or only marginally ineffective in most systems, with the possible exceptions of V455 And, XY Ari and GK Per. 

In the following, we have taken $f=1$ and $\omega_*=0$, which maximizes the effect of changing the inner boundary condition; the steady state solution now becomes $\nu\Sigma = \dot{M}/3\pi$, in which the $[1-(r_{\rm in}/r)^{1/2}]$ term that is present in the standard case has disappeared.

As can be seen from Table \ref{tab:results}, changing the boundary condition has a significant, but not major, effect. It destabilizes the disc, and is somehow equivalent in decreasing the magnetic moment.

\section{Enhanced mass-transfer rate}

The DIM does not appear to be able to produce very short duration (less than a day) outbursts, in particular in systems which have long orbital periods. It is worth therefore to study the effect  that mass transfer variations can have on magnetically truncated discs in IPs. Such variations on various time-scales are universally observed in cataclysmic variables, such as AM Her stars, and have been invoked to account for some specific systems such as V513 Cas or IW And, two Z Cam systems which exhibited dwarf nova outbursts during their standstill before entering into a low state \citep{hl14}. These outbursts cannot be explained by the DIM since the disc in Z Cam systems during standstills is assumed to be on the hot branch. \citet{hb93} suggested that such mass transfer outbursts were responsible for the short outbursts observed in TV Col, based on the fact that the emission lines, presumably originating from the hot spot region as indicated by the trailed spectrograms, were increased by a factor 30 as compared to quiescence. It is therefore worth exploring the possibility that these short outbursts could be explained by enhanced mass-transfer rate. Even for SS Cyg one has to take into account mass-transfer variations to obtain good agreement of models with observations \citep{SHL03}.

   \begin{figure}
   \centering
   \includegraphics[angle=-90,width=\columnwidth]{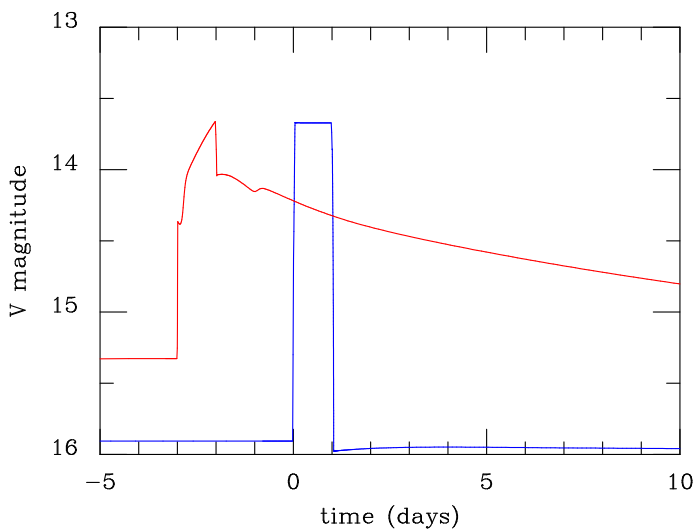}
   \caption{Visual magnitude of a system undergoing a mass transfer outburst (mass transfer rate multiplied by 20), for initial $\dot{M}_{\rm tr} = 10^{18}$ gs$^{-1}$ (red curve) and $10^{16}$ gs$^{-1}$ (blue curve). The magnitudes have been scaled to be comparable. The outbursts starts at $t=-3$ (red curve) and $t=0$ (blue curve).}
   \label{fig:mtr}
   \end{figure}
   
One requires, however, that the disc either lies on the cold stable branch, or on the hot stable branch, since these the systems experiencing short outbursts do not show normal, longer, dwarf nova outbursts. We therefore consider both possibilities. Figure \ref{fig:mtr} hows the evolution of a systems with an orbital period of 6 hr in which mass transfer is increased by a factor 20 during 1 day. For the high mass transfer case, $\dot{M}_{\rm tr} = 10^{18}$ g s$^{-1}$, and the magnetic moment is $2 \times 10^{33}$ Gcm$^3$; for the low mass transfer case, $\dot{M}_{\rm tr} = 10^{16}$ gs$^{-1}$, and the magnetic moment is $2 \times 10^{32}$ Gcm$^3$. $\dot{M}_{\rm tr}$ is close to the minimum  (resp. maximum) mass transfer possible for a steady state on the hot (resp. cold) branch of the S-curve, and the magnetic moment has been chosen high so as to minimize the disc mass. In both cases, the contribution of the hot spot, whose temperature is 15,000K, has been included; $M_1$ is taken to be 0.8 M$_\odot$, and $q=0.6$. In both cases, the amplitude of the outburst is about 2 magnitudes, compatible with the observations. 

In the high $\dot{M}_{\rm tr}$ case, the disc mass is comparable to the total mass transferred during an outburst, and, because it lies on the hot stable branch, the viscous time is relatively short and the disc response is significant and occurs on a time scale of a few days. As a consequence, the outburst is very asymmetric: sharp rise, due to the immediate increase of the hot spot luminosity, immediately followed by the disc response, and a slow decline, on the disc viscous time scale after a sharp decline of the hot spot luminosity; the peak of the mass accretion rate onto the white dwarf occurs 1.7 days after the beginning of the mass transfer event, and is $5.5 \times 10^{18}$ gs$^{-1}$. In the low $\dot{M}_{\rm tr}$ case, the disc mass is more than two orders of magnitude larger than the mass transferred during the outburst, because of the much reduced mass transfer rate and of the lower viscosity; as a consequence, the disc is not much affected by the mass transfer and responds on a much longer viscous time scale: the mass accretion rate onto the white dwarf reaches its maximum 330 days after the beginning of the outburst, and the relative increase is only of about 4\%.

A sudden increase of the mass transfer rate from the secondary could therefore in principle explain the optical light curve of the short outbursts observed in intermediate polars, provided that the quiescent mass transfer rate is low enough to allow it to remain cold and stable. This increase is not sufficient to trigger a standard dwarf nova outburst, and the accretion rate onto the white dwarf cannot increase by more than few percent. This conclusion relies, however, on the assumption that the whole mass-transfer stream is stopped at the outer disc edge, which is probably not correct, since it is very likely that the stream overflowing the disc interacts directly with the magnetosphere, as suggested by e.g. \citet{hkn00} in the case of EX Hya. This is anyway required to account for the X-ray emission accompanying the optical outbursts in systems such as EX Hys \citep{hkn00}, YY Dra \citep{sne02}.

\section{Conclusions}

We have shown that the white dwarf magnetic field can stabilize the accretion disc on the cold branch in intermediate polar systems, explaining the existence of quiescent intermediate polars, in particular below the period gap where all non magnetic, H-rich systems, are observed to be dwarf novae with very few exceptions that can be accounted for by some specificity of the system, such as an historical nova explosion. The DIM can account for the observed outbursts properties of systems which undergo (normal) long outbursts, such as XY Ari or YY Dra. The DIM cannot explain the very short, rare, outbursts observed in a number of systems. Our results contradict the early ones from \citet{av89} who assumed high viscosities which, we now know, are unrealistic.

These outbursts could, in principle, be due to the mass-transfer rate from the secondary enhanced by a factor $\sim$ 20, provided that the quiescent mass transfer rate is low enough that the disc can sit on the cold stable branch. This hypothesis gains support from the fact that high excitation emission lines are strongly enhanced during an outburst, which, from their time-dependant velocities, seem to originate from the hot spot. In order to account for the concomitant X-ray outburst, the stream from the $L_1$ point must partly overflow the accretion disc and interact directly with the white dwarf magnetosphere. This, however, lowers the efficiency of optical emission, and requires an even larger increase of the mass transfer rate. It is also unclear that the hot spot can emit the observed very high excitation lines; we have assumed here a hot spot temperature of 15,000 K, which maximizes its optical emission, but is most probably too low so that $\dot{M}_{\rm tr}$ probably needs to be increased even more.

For all these reasons, an enhanced mass-transfer outburst, although possible, is not necessarily the explanation of short dwarf nova outbursts in IPs. Such large-amplitude, short outbursts should also be present in non magnetic systems -- and they might be present, see for example the case of V1316 Cyg \citep{sbp06} which, in addition to standard dwarf nova outbursts, exhibits short (less than 1-2 days), faint (average amplitude of 1.4 magnitude), relatively frequent (6 outbursts recorded over 4 months in 2005) events. The suggestion by \citet{mbd07} that EX Hya outbursts could be due to the storage and release of matter outside the magnetosphere, with a similar mechanism to that put forward by \citet{st93} to explain the \textsl{Rapid Burster}, is interesting, but it remains to be explained why the recurrence time between short DN outbursts in IPs is so long, longer than the recurrence time. Another similar possibility is that the coupling of the magnetic field generated by the MRI with the magnetic field of the white dwarf can, under particular conditions, generate an instability. Exploring this option requires global numerical simulations of the MRI, which are now beginning to be possible  \citep[see e.g.][]{zs17}; the inclusion of the white dwarf magnetic field remains, however, to be implemented.

\begin{acknowledgements}
This work benefited from many discussions with the participants of the KITP 2017 program "Confronting MHD Theories of Accretion Disks with Observations", in particular with Paula Szkody and Christian Knigge. This work was supported by a National Science Centre, Poland grant 2015/19/B/ST9/01099 and by the National Science Foundation under Grant No. NSF PHY-1125915. JPL was supported by a grant from the French Space Agency CNES.
\end{acknowledgements}


\begin{thebibliography}{99}

\bibitem[Andronov et al.(2008)]{ach08} Andronov, I.~L., Chinarova, L.~L., Han, W., Kim, Y., \& Yoon, J.-N. 2008, A\&A, 486, 855 
\bibitem[Angelini \& Verbunt(1989)]{av89}Angelini, L., \& Verbunt, F. 1989, MNRAS, 238, 697
\bibitem[Buat-M\'enard et al.(2001)]{bhl01}Buat-Ménard, V., Hameury, J.-M., \&  Lasota, J.-P. 2001, A\&A, 366, 612
\bibitem[Crawford et al.(2008)]{cbg08} Crawford, T., Boyd, D., Gualdoni, C., et al. 2008, JAVSO, 36, 60 
\bibitem[Duschl \& Tscharnuter(1991)]{dt91}Duschl, W.~J., \& Tscharnuter, W.~M. 1991, A\&A, 241, 153
\bibitem[Frank et al.(2002)]{FKR02} Frank, J., King, A., \& Raine, D.~J.\ 2002, Accretion Power in Astrophysics, by Juhan Frank and Andrew King and Derek Raine, pp.~398.~ISBN 0521620538.~Cambridge, UK: Cambridge University Press, February 2002., 398 
\bibitem[Gaia Collaboration et al.(2016)]{gaia} Gaia Collaboration, et al., 2016, A\&A, 595, A2
\bibitem[Hameury \& Lasota(2002)]{hl02}Hameury, J.-M., \& Lasota, J.-P. 2002, A\&A, 394, 231
\bibitem[Hameury \& Lasota(2014)]{hl14}Hameury, J.-M., \& Lasota, J.-P. 2014, A\&A, 569, A48
\bibitem[Hameury et al.(1986)]{hkl86}Hameury, J.-M., King, A.~R., \& Lasota, J.-P. 1986, MNRAS, 218, 695
\bibitem[{{Hameury} {et~al.}(1998){Hameury}, {Menou}, {Dubus}, {Lasota}, \&
      {Hure}}]{hmdlh98}{Hameury}, J-M.., {Menou}, K., {Dubus}, G., {Lasota}, J.-P., \& {Hure}, J.-M. 1998, MNRAS, 298, 1048
\bibitem[Hameury et al.(2000)]{hlw00}Hameury, J.M., Lasota, J.P., \& Warner, B. 2000, A\&A, 353, 244
\bibitem[Harrison et al.(1999)]{hms99}Harrison, T.~E., McNamara, B.~J., Szkody, P., et al. 1999, ApJ, 515, L93
\bibitem[Hellier \& Buckley(1993)]{hb93}Hellier, C., \& Buckley, D.~A.~H. 1993, MNRAS, 265, 766
\bibitem[Hellier et al.(1997)]{hmb97} Hellier, C., Mukai, K., \& Beardmore A.~P. 1997, MNRAS, 292, 397
\bibitem[Hellier et al.(2000)]{hkn00} Hellier, C., Kemp, J., Naylor, T., et al.\ 2000, \mnras, 313, 703 
\bibitem[Hudec et al.(2005)]{hss05}Hudec, R., Šimon, V., \& Skalický, J. 2005, ASPC, 330, 405
\bibitem[Ishioka et al.(2002)]{i02} Ishioka, R., Kato, T., Uemura, M., et al. 2002, PASJ, 54, 581 
\bibitem[Johnson et al.(2017)]{jth17} Johnson, C.~B., Torres, M.~A.~P., Hynes, R.~I., et al. 2017, MNRAS, 466, 129 
\bibitem[Kato et al.(2009)]{kiu09} Kato T., Imada, A., Uemura, M., et al. 2009, PASJ, 61, S395 
\bibitem[Kato et al.(2015)]{kho15}Kato, T., Hambsch, F.-J., Oksanen, A., Starr, P., \& Henden, A. 2015, PASJ, 67, 3 
\bibitem[Kim et al.(1992)]{kwm92} Kim S.-W., Wheeler J.~C., \& Mineshige S. 1992, ApJ, 384, 269 
\bibitem[Knigge et al.(2011)]{kbp11}Knigge, C., Baraffe, I., \& Patterson, J. 2011, ApJS, 194, 28
\bibitem[Kotko \& Lasota(2012)]{KL12} Kotko, I., \& Lasota, J.-P.\ 2012, \aap, 545, A115 
\bibitem[Kotko et al.(2012)]{KLDH12} Kotko, I., Lasota, J.-P., Dubus, G., \& Hameury, J.-M.\ 2012, \aap, 544, A13 
\bibitem[Lasota(2001)]{l01}{Lasota}, J.-P. 2001, \nar, 45, 449
\bibitem[Lasota et al.(1995)]{lhh95} Lasota J.-P., Hameury J.-M.,\&  Hure J.-M. 1995, A\&A, 302, L29
\bibitem[Lasota \& Schreiber(2007)]{ls07}Lasota, J.-P.., Schreiber, M.~R. 2007, A\&A, 473, 897
\bibitem[Littlefield et al.(2016)]{lgk16}Littlefield, C.,  Garnavich, P., Kennedy, M.~R., et al. 2016, ApJ, 833, 93
\bibitem[Lubow \& Shu(1975)]{ls75}Lubow, S.~H., \& Shu, F.~H. 1975, ApJ, 198, 383
\bibitem[Mason at al.(2013)]{mom13}Mason, E., Orio, M., Mukai, K., et al. 2013, MNRAS, 436, 212
\bibitem[Mhlahlo et al.(2007)]{mbd07} Mhlahlo, N., Buckley, D.~A.~H., Dhillon, V.~S., et al.\ 2007, \mnras, 380, 353 
\bibitem[Miller-Jones et al.(2013)]{msk13}Miller-Jones J.~C.~A., Sivakoff G.~R., Knigge C., et al., 2013, Sci, 340, 950
\bibitem[Nogami et al.(2009)]{nhz09}Nogami, D., Hiroi, K., Suzuki, Y., et al. 2009, ASPC, 404, 52
\bibitem[Paczy{\'n}ski(2000)]{BP00} Paczy{\'n}ski, B.\ 2000, arXiv:astro-ph/0004129 
\bibitem[Patterson et al.(2013)]{puk13}Patterson, J., Uthas, H., Kemp, J., et al. 2013, MNRAS, 434, 1092
\bibitem[Ritter \& Kolb(2003)]{rk03} Ritter, H., \& Kolb U. 2003, A\&A, 404, 301 
\bibitem[Rude \& Ringwald(2012)]{rr12} Rude, G.~D., \& Ringwald, F.~A. 2012, NewA, 17, 442 
\bibitem[Shakura \& Sunyaev(1973)]{ss73}Shakura, N.~I. \& Sunyaev, R.~A. 1973, A\&A, 24, 337
\bibitem[Schreiber \& Gänsicke(2002)]{sh02}Schreiber, M.~R., \& Gänsicke, B.~T. 2002, A\&A, 382, 124
\bibitem[Schreiber et al.(2003)]{SHL03} Schreiber, M.~R., Hameury, J.-M., \& Lasota, J.-P.\ 2003, \aap, 410, 239 
\bibitem[Shears et al.(2006)]{sbp06} Shears, J., Boyd, D., \& Poyner, G.\ 2006, Journal of the British Astronomical Association, 116, 244 
\bibitem[Šimon(2015)]{s15}Šimon, V. 2015, A\&A, 575, A65
\bibitem[Smak(1983)]{Smak83} Smak, J.\ 1983, \actaa, 33, 333 
\bibitem[Smak(1999)]{Smak99} Smak, J.\ 1999, \actaa, 49, 391 
\bibitem[Spruit \& Taam(1993)]{st93} Spruit, H.~C., \& Taam, R.~E.\ 1993, \apj, 402, 593 
\bibitem[Szkody \& Mateo(1984)]{sm84}Szkody, P., \& Mateo, M. 1984, ApJ, 280, 729
\bibitem[Szkody et al.(2002)]{sne02} Szkody, P., Nishikida, K., Erb, D., et al.\ 2002, \aj, 123, 413
\bibitem[van Amerongen \& van Paradijs(1989)]{vv89}van Amerongen, S., \& van Paradijs, J. 1989, A\&A, 219, 195
\bibitem[Viallet \& Hameury(2007)]{vh07}Viallet, M., \& Hameury, J.-M. 2007, A\&A, 475, 597
\bibitem[Viallet \& Hameury(2008)]{vh08}Viallet, M., \& Hameury, J.-M. 2007, A\&A, 489, 699
\bibitem[Wagner et al.(1998)]{wth98}Wagner, R.~M., Thorstensen, J.~R., Honeycutt, R.~K., et al. 1998, AJ, 115, 787
\bibitem[Warner(1995)]{w95}Warner, B. 1995, Cataclysmic variable stars, Camb. Astrophys. Ser., 28
\bibitem[Warner et al.(1996)]{wlt96} Warner B., Livio M., \& Tout C.~A. 1996, MNRAS, 282, 735
\bibitem[Zhu \& Stone(2017)]{zs17} Zhu, Z., \& Stone, J.~M.\ 2017, arXiv:1701.04627 
\end{thebibliography}
\end{document}